\documentclass[12pt]{article}

\usepackage{scicite}

\usepackage{mathptmx}

\newenvironment{sciabstract}{%
\begin{quote} \bf}
{\end{quote}}

\usepackage{graphicx,amssymb,amsmath}
\newcommand {\bisco}{Bi$_2$Sr$_2$CaCu$_2$O$_{8+\delta}$}
\newcommand {\uJcm}{$\mu$J/cm$^2$}

\newcounter{lastnote}
\newenvironment{scilastnote}{%
\setcounter{lastnote}{\value{enumiv}}%
\addtocounter{lastnote}{+1}%
\begin{list}%
{\arabic{lastnote}.}
{\setlength{\leftmargin}{.22in}}
{\setlength{\labelsep}{.5em}}}
{\end{list}}

\title{Tracking Cooper Pairs in a Cuprate Superconductor by Ultrafast Angle-Resolved Photoemission}

\author
{Christopher L. Smallwood,$^{1,2}$ James P. Hinton,$^{1,2}$ Christopher Jozwiak,$^{3}$\\ 
Wentao Zhang,$^{2}$ Jake D. Koralek,$^{2}$ Hiroshi Eisaki,$^{4}$ Dung-Hai Lee,$^{1,2}$\\
Joseph Orenstein,$^{1,2}$ Alessandra Lanzara$^{1,2\ast}$\\
\\
\normalsize{$^{1}$Department of Physics, University of California,}\\
\normalsize{Berkeley, CA 94720, USA}\\
\normalsize{$^{2}$Materials Sciences Division, Lawrence Berkeley National Laboratory,}\\
\normalsize{Berkeley, CA 94720, USA}\\
\normalsize{$^{3}$Advanced Light Source, Lawrence Berkeley National Laboratory,}\\
\normalsize{Berkeley, CA 94720, USA}\\
\normalsize{$^{4}$Electronics and Photonics Research Institute,}\\
\normalsize{National Institute of Advanced Industrial Science and Technology,}\\
\normalsize{Ibaraki 305-8568, Japan}\\
\\
\normalsize{$^\ast$To whom correspondence should be addressed; E-mail: alanzara@lbl.gov.}
}

\date{}

\begin{document} 




\maketitle


\begin{sciabstract}
In high-temperature superconductivity, the process that leads to the formation of Cooper pairs, the fundamental charge carriers in any superconductor, remains mysterious.
We use a femtosecond laser pump pulse to perturb superconducting \bisco, and study subsequent dynamics using time- and angle-resolved photoemission and infrared reflectivity probes. 
Gap and quasiparticle population dynamics reveal marked dependencies on both excitation density and crystal momentum. 
Close to the $d$-wave nodes, the superconducting gap is sensitive to the pump intensity and Cooper pairs recombine slowly.
Far from the nodes pumping affects the gap only weakly and recombination processes are faster. 
These results demonstrate a new window into the dynamical processes that govern quasiparticle recombination and gap formation in cuprates.
\end{sciabstract}


The lifetime of Bogoliubov quasiparticles, the low energy excitations of a superconductor, contains a wealth of information pertinent to the origin of superconductivity in a given material\cite{Schrieffer99}. 
This lifetime reflects two distinct processes: quasiparticle scattering and recombination.
In the former, a quasiparticle scatters from one momentum state to another, conserving the fermionic particle number.
Recombination, on the other hand, refers to interactions in which two quasiparticles annihilate.
To conserve energy and momentum, recombination must involve emission of other excitations, for example phonons or magnons, to which the quasiparticles are strongly coupled.
Measurement of quasiparticle recombination rates as a function of their energy and momentum can, in principle, provide direct information about the interactions that induce Cooper pairing and superconductivity\cite{Kaplan76}.
Only very recently, with the demonstration that angle-resolved photoemission spectroscopy (ARPES) can be performed with ultrashort laser pulse sources\cite{Perfetti07,Schmitt08,Graf11,Cortes11,Rohwer11,Petersen11,Rettig12}, have measurements with the necessary energy, momentum, and time resolution become possible. 
We present the results of experiments that use synchronized laser pulses to perform time-resolved ARPES measurements of quasiparticle recombination and gap dynamics in the high-temperature superconductor \bisco.

Measurements were performed at 18 K on an optimally doped sample with a critical temperature ($T_c$) of 91 K.
A transient state is created with an infrared laser pump pulse ($h\nu = 1.48$ eV) and measured via photoemission shortly thereafter, with a temporal resolution of 300 fs, using an ultraviolet probe pulse ($h\nu = 5.9$ eV).
The experiment benefits from high momentum and energy resolution (0.003 \AA$^{-1}$ and 23 meV, respectively) and the ability to explore low pump fluences (2-15 \uJcm).

Figure 1 shows typical equilibrium and transient ARPES dispersions ($t=-1$ ps and $t=0.6$ ps respectively) for cuts along nodal and off-nodal directions in $k$-space. 
\begin{figure}[htbp]\centering\includegraphics[width=0.9\columnwidth]{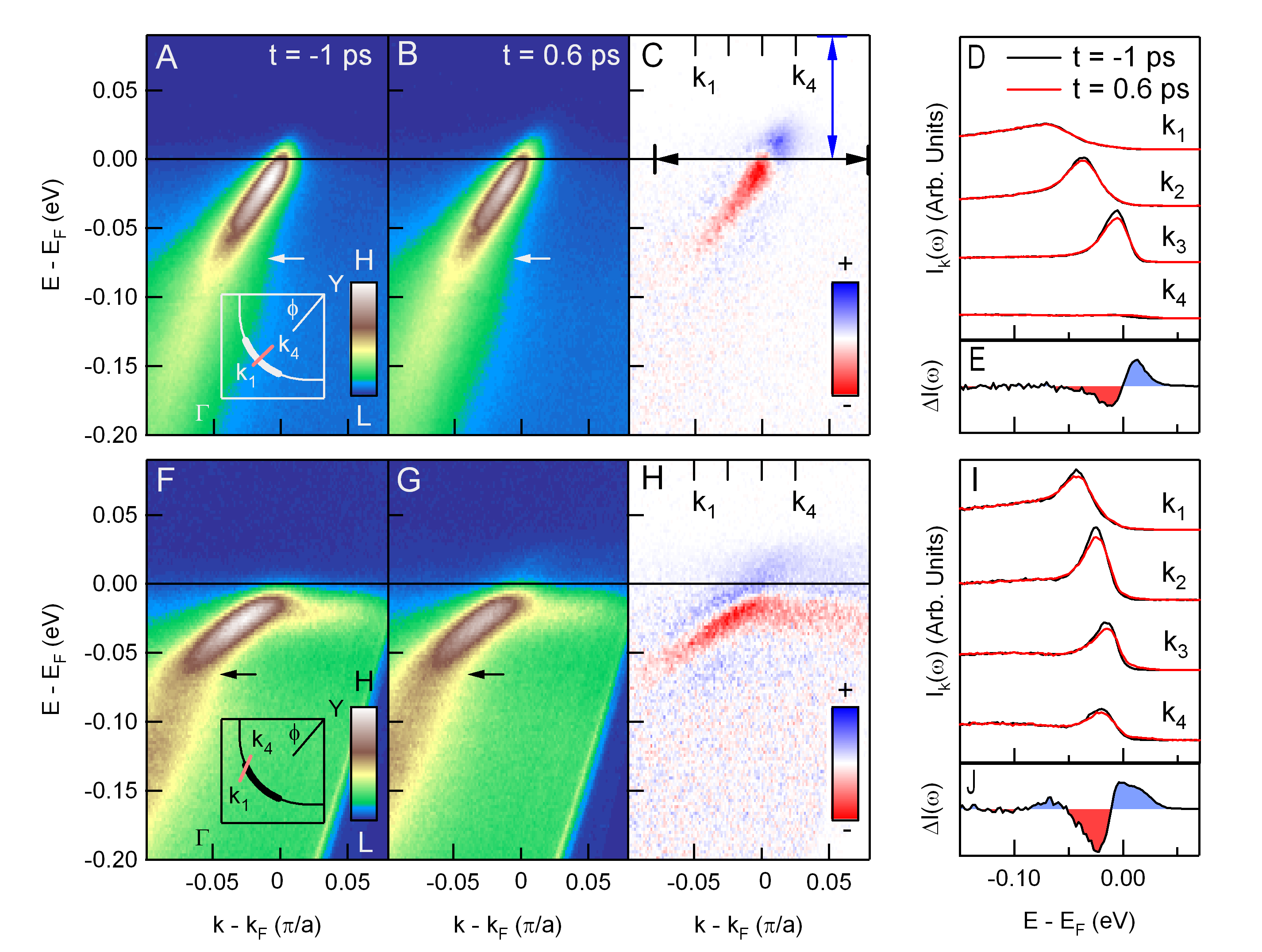}
\caption{\label{fig1}
Typical ARPES dispersions before and after pumping for nodal ($\phi=45^\circ$) and gapped ($\phi=31^\circ$) regions  of $k$-space. 
The incident pump fluence was 5 \uJcm. 
(A) Equilibrium ($t = -1$ ps) and (B) transient ($t = 0.6$ ps) energy-momentum maps for the nodal state. Data are shown with identical color scales. The inset shows the location of the cut. The arrow marks the position of the dispersion kink. 
(C) Subtraction between (A) and (B). Blue indicates intensity gain and red intensity loss. 
(D) Energy distribution curves (EDCs) from ${k_1}$ to ${k_4}$ for equilibrium (in black) and transient (in red) states. EDCs are shifted vertically for ease of comparison.
(E) Difference between transient and equilibrium EDCs, integrated across the double black arrow in panel (C).
(F-J) Same as (A-E) but for a gapped (off-nodal) momentum cut.
Spectra have been corrected for detector non-linearity. The diagonal line in the lower right portion of (F) and (G) is the edge of the detector.
}
\end{figure}
Here the time origin $t=0$ coincides with the application of the pump pulse.
The off-nodal cut has an equilibrium gap of 15 meV.
In both cuts a well-defined kink (marked by arrows in panels (A-B) and (F-G))\cite{Lanzara01,Cuk04} separates sharply defined coherent dispersive features from poorly defined incoherent features, as also visible in the selected energy distribution curves (EDCs) shown in panels (D) and (I).
The following changes are evident in the transient spectra:
1) a decrease of intensity below the Fermi level ($E_F$) and slight broadening in the coherent spectra (panels (C-E), and (H-J)), similar to a previous report for nodal quasiparticles\cite{Graf11} and mainly confined below the kink binding energies\cite{Lanzara01,Cuk04};
2) an overall transfer of spectral weight across $E_F$ (panels (C), (E), (H), and (J)), indicating the creation of transient quasiparticles; 
and 3) a small shift of the spectral peak toward $E_F$ in the off-nodal cut (panels (H) and (I)), indicating a partial closure of the superconducting gap.

Figure 2 shows the temporal evolution of the superconducting gap in response to photoexcitation, as extracted from symmetrized EDCs at $k_F$, the Fermi wave vector\cite{Norman98a}. 
\begin{figure}\centering\includegraphics[width=0.8\columnwidth]{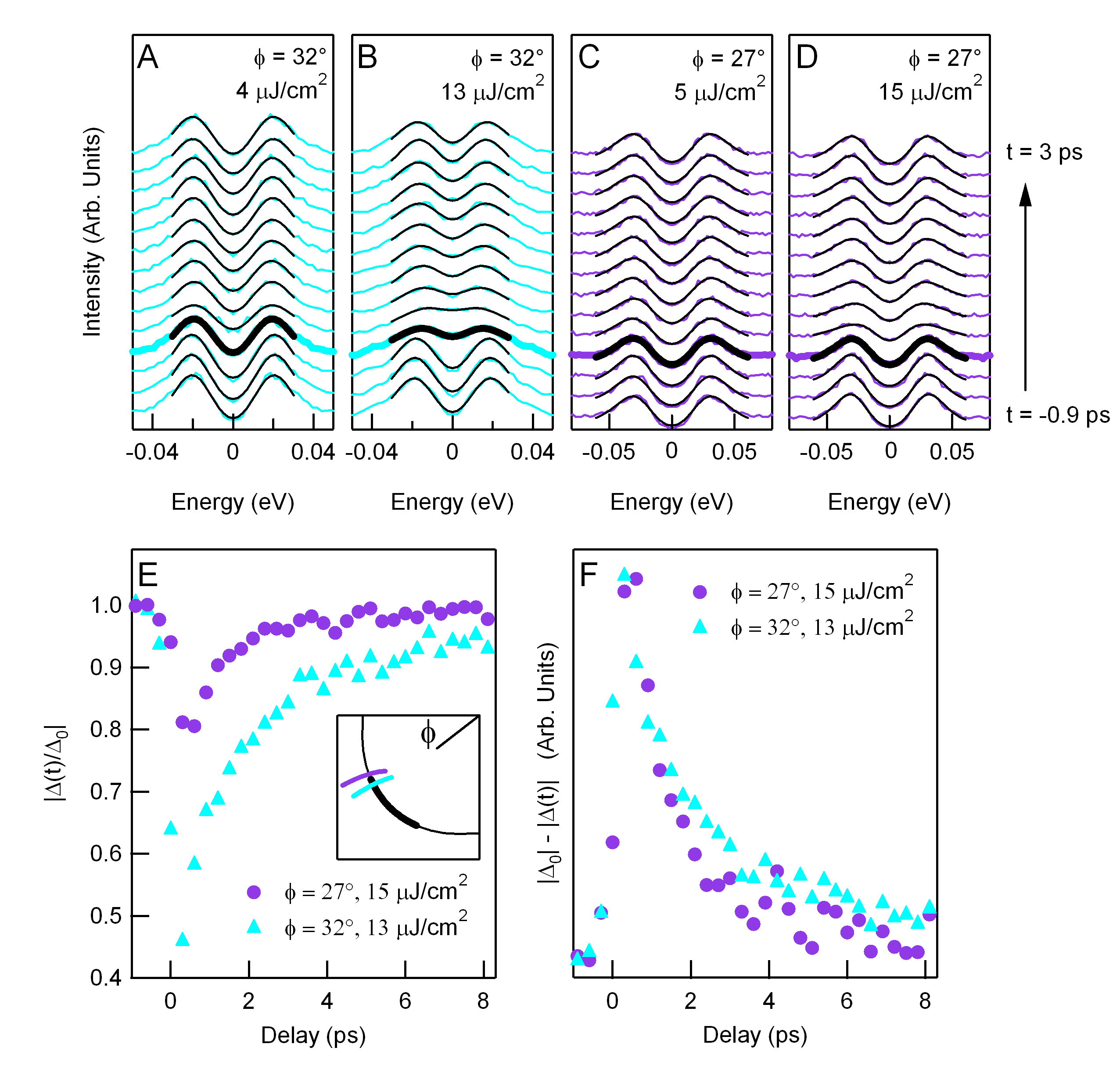}
\caption{\label{gap}
Evolution of the superconducting gap following pump excitation.
Symmetrized EDCs at $k_F$ for $\phi=32^\circ$ at low (A) and higher (B) fluence. The gap is obtained by fitting to a phenomenological model\cite{Norman98a}, but can be approximated by halving the distance between positive and negative peaks. Bold curves correspond to $t=0$. 
For additional gap fitting details, see supplementary information online\cite{som}.
(C-D) Analogous EDCs for a cut at $\phi=27^\circ$.
(E) Gap magnitude normalized by its equilibrium value vs.\ pump-probe delay for momentum cuts at $\phi=27^\circ$ and $\phi=32^\circ$. 
(F) Gap magnitude, inverted and normalized by maximal change upon pumping in order to compare recovery rates.
}
\end{figure}
Panels (A-B) and (C-D) correspond to two representative cuts at $\phi=32^\circ$ and $\phi=27^\circ$, respectively, with $\phi$ defined according to the inset of panel (E). 
These spectra indicate very different responses of the gap amplitude to photoexcitation for the two cuts.
The gap is relatively insensitive to fluences below 5 \uJcm\, but 13 \uJcm\ induces a clear reduction in size. 
As shown in panel (E) the gap closer to the node decreases by 55\% of its equilibrium magnitude, while the gap at $\phi=27^\circ$ decreases by only 20\%. 
This may indicate different dynamics inside and outside the Fermi arc, which is reported to end rather abruptly at $\phi=30^\circ$ for samples of this doping\cite{Tanaka06,Lee07}, although studies farther from the node are needed.
Gap recovery rates are illustrated in Fig.\ 2F, where the curves from panel (E) have been inverted and rescaled by their maximum change.
The initial recovery rate is slower for states closer to the node ($0.9 \pm 0.6$ ps$^{-1}$) than for states farther from the node ($1.3 \pm 0.6$ ps$^{-1}$), although the contrast is less apparent than that between amplitudes.

Figs.\ 3 and 4 show quasiparticle recombination dynamics.  
\begin{figure}[htbp]\centering\includegraphics[width=0.9\columnwidth]{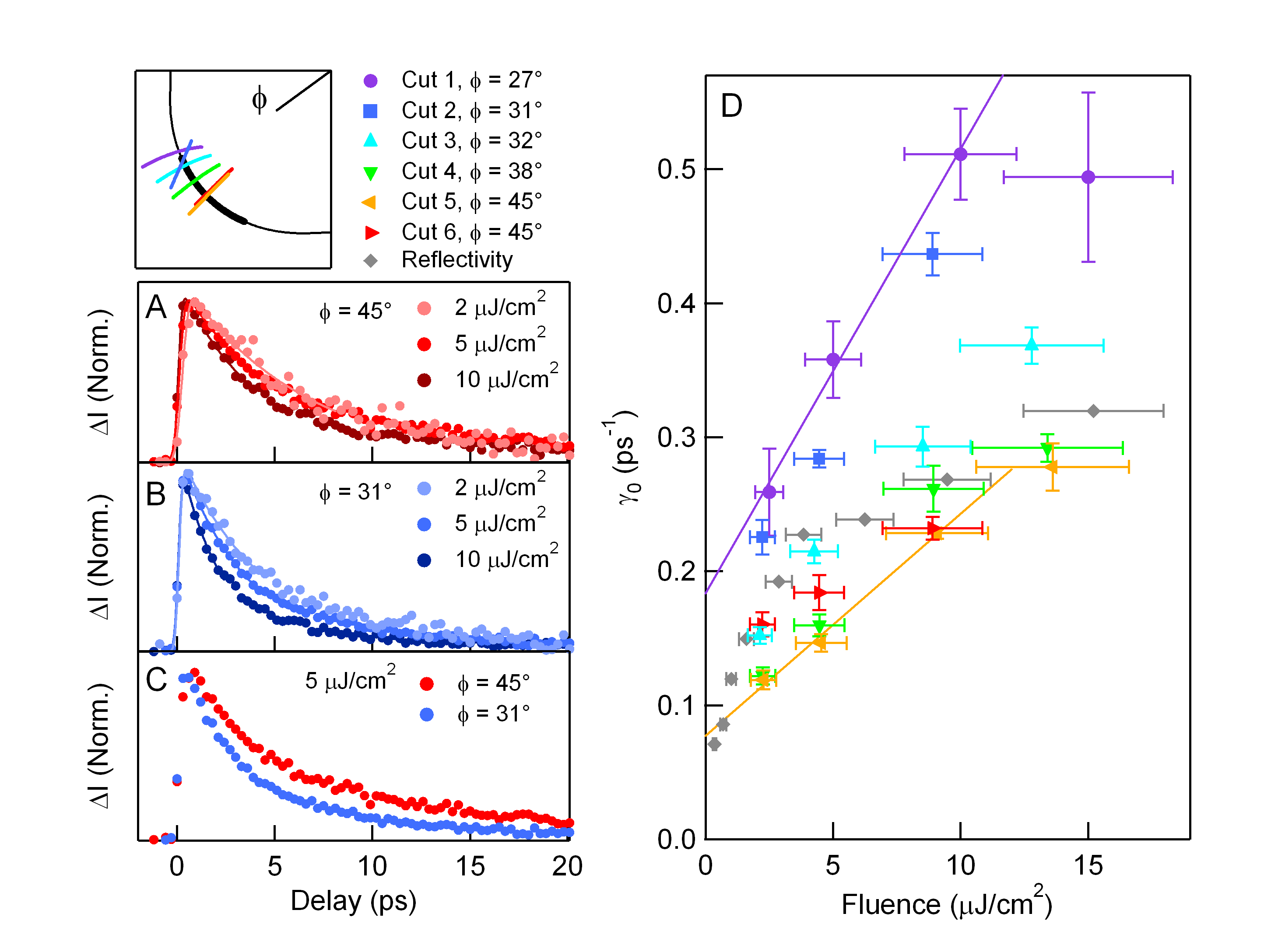}
\caption{\label{excurves}
Quasiparticle recombination dynamics vs.\ pump fluence and crystal momentum. 
ARPES data correspond to intensity change above $E_F$ ($\Delta I$) as integrated between the blue and black double arrows in Fig.\ 1C. 
Time-resolved reflectivity rates correspond to fractional change in reflectivity. 
(A) Nodal decay curves at 2, 5, and 10 \uJcm, normalized to the same amplitude. (B) Analogous off-nodal decay curves ($\phi=31^\circ$). 
(C) Overlay of nodal and off-nodal curves at the same fluence. 
(D) Initial decay rate $\gamma_0$ vs.\ fluence, obtained by fitting decay curves at short times (for $\Delta I (t) \gtrsim \Delta I_0/2$) to the convolution of a Gaussian and the function $f(t)=\Delta I_0 e^{-\gamma_0(t-t_0)} \Theta(t-t_0)$, where 
$\Delta I_0$ and $t_0$ are additional fit parameters. 
Time-resolved reflectivity rates were multiplied by 3/2 in order to take the finite penetration depth of the optical probe into account\cite{som}.
}
\end{figure}
In this low fluence regime the gap is almost unchanged for most of the recovery process, so quasiparticle recombination is largely decoupled from gap dynamics.
Fig.\ 3A-C shows the temporal evolution of the above-$E_F$ spectral change $\Delta I$ for representative nodal and off-nodal $k$-space cuts, where $\Delta I$ is defined by the integrated intensity change across the blue and black double arrows in Fig.\ 1C. 
The spectral change is nearly symmetric above and below $E_F$ in this fluence regime, so we focus on the intensity above $E_F$ because of its superior statistics and smaller background.
Faster decay rates occur at higher fluences and off-nodal momenta, an effect which cannot be explained by equilibrium heating\cite{som}.

Figure 3D summarizes the dependence of quasiparticle recombination on fluence and momentum.
The rate $\gamma_0$ is defined by fitting the decay curves at short times to the convolution of a Gaussian and decaying exponential\cite{Segre02,Gedik04}.
In line with Fig.\ 3A-C, two prominent decay rate trends are apparent: 
1) fluence dependence, with faster initial decay rates $\gamma_0$ occurring at higher fluences; 
and 2) momentum dependence, with off-nodal decay rates increasing faster with fluence than those at the node.
The first trend implies that intrinsic quasiparticle recombination processes are observed\cite{Gedik04,Kaindl05}.
The second trend indicates that this recombination occurs more rapidly in off-nodal regions of $k$-space than at the node. 
The fluence dependence is also complementary to ultrafast studies using all-optical techniques, which report a dramatic decay rate fluence dependence, particularly in the low fluence regime\cite{Segre02,Gedik04,Gedik05,Kaindl05,Coslovich11}, and there is overall agreement between the ARPES results and a time-resolved reflectivity measurement taken on the same sample (gray circles in Fig.\ 3D).
The decay rate measured by reflectivity is uniformly faster than nodal ARPES decay rates (cuts 5-6), but slower than the off-nodal rates (cuts 1-3), suggesting that optical spectroscopy provides an effective momentum-integrated average of the quasiparticle population. 
Interestingly, along both $\phi=27^\circ$ and $\phi=32^\circ$ directions quasiparticle decay rates are slower than the gap recovery rates in Fig.\ 2: compare $1.3 \pm 0.4$ ps$^{-1}$ and $0.9 \pm 0.6$ ps$^{-1}$ for the gap recovery vs.\ $0.49 \pm 0.06$ ps$^{-1}$ and $0.37 \pm 0.01$ ps$^{-1}$ for the intensity recovery.  
This indicates that the superconducting gap recovers well before the nonequilibrium quasiparticle population drops to zero. 
A potentially related effect occurs at equilibrium in the BCS model, where the gap becomes large for $T$ only slightly below $T_c$.
Finally, we note that the fluence and momentum dependencies reported here are in contrast to the findings of another time-resolved ARPES work\cite{Cortes11}, where the quasiparticle recombination rate was reported to be independent of both fluence and momentum.
The discrepancy might be explained by the higher fluences used in the previous work, which likely result in a complete closure of the superconducting gap, or by the coarser momentum and energy resolution compared to the present study.  

As noted above (see also supplementary information online\cite{som}), the fluence dependence of 
$\gamma_0$ means that ARPES decay rates are connected to intrinsic quasiparticle recombination processes.
Time-resolved optical measurements indicate that the total (momentum-integrated) population of photoexcited quasiparticles $n_{ex}(t)$ can be described by a bimolecular rate equation\cite{Gedik04,Kaindl05},
\begin{equation}
\frac{\dot{n}_{ex}}{n_{ex}} = -R \left( n_{ex} + 2 n_{T}  \right), \label{bimolecular}
\end{equation}
where $R$ is a quasiparticle recombination constant, and $n_T$ is the population of thermal quasiparticles.
This is a special case of the Rothwarf-Taylor model of quasiparticle recombination\cite{Rothwarf67}, which has been successfully used to model dynamics of both conventional and high-temperature superconductors\cite{Demsar03,Gedik04,Gedik05,Kaindl05,Kabanov05,Kusar08,Giannetti09,Coslovich11,Torchinsky10,Torchinsky11,Beck11}. 
The general model also incorporates negative feedback from ``hot'' bosons ($\hbar \omega \gtrsim 2\Delta$) that are created by the quasiparticle decay process. For low fluence and temperature the model reduces to Eq.\ \ref{bimolecular} under both weak feedback and strong feedback (boson bottleneck) scenarios, which beget differing interpretations of the coefficient $R$\cite{Gedik04,Kabanov05}. 
In both approximations bimolecular recombination is the active ingredient in fluence-dependent dynamics.

In contrast to time-resolved optics, ARPES measures the momentum-dependent nonequilibrium quasiparticle density $n_k(t)$. 
The short-time fluence-dependent recombination dynamics are given by
\begin{equation}
\frac{\dot{n}_k}{n_k} \approx - \int R_{kk'} \, n_{k'} \, d^2k', \label{bimolecular2}
\end{equation}
where $R_{kk'}$ is a modified recombination coefficient for the interaction between quasiparticles at specific points $k$ and $k'$ in reciprocal space. A weighted average of $R_{kk'}$ over $k'$ is given by $\partial \gamma_0/\partial F$, the rate of increase of the initial decay rate $\gamma_0$ with fluence\cite{som}.
Fig.\ 4 shows an analysis of $\partial \gamma_0/\partial F$ as calculated by fitting straight lines to the data in Fig.\ 3D for fluences $F < 12$ \uJcm.
\begin{figure}[htbp]\centering\includegraphics[width=\columnwidth]{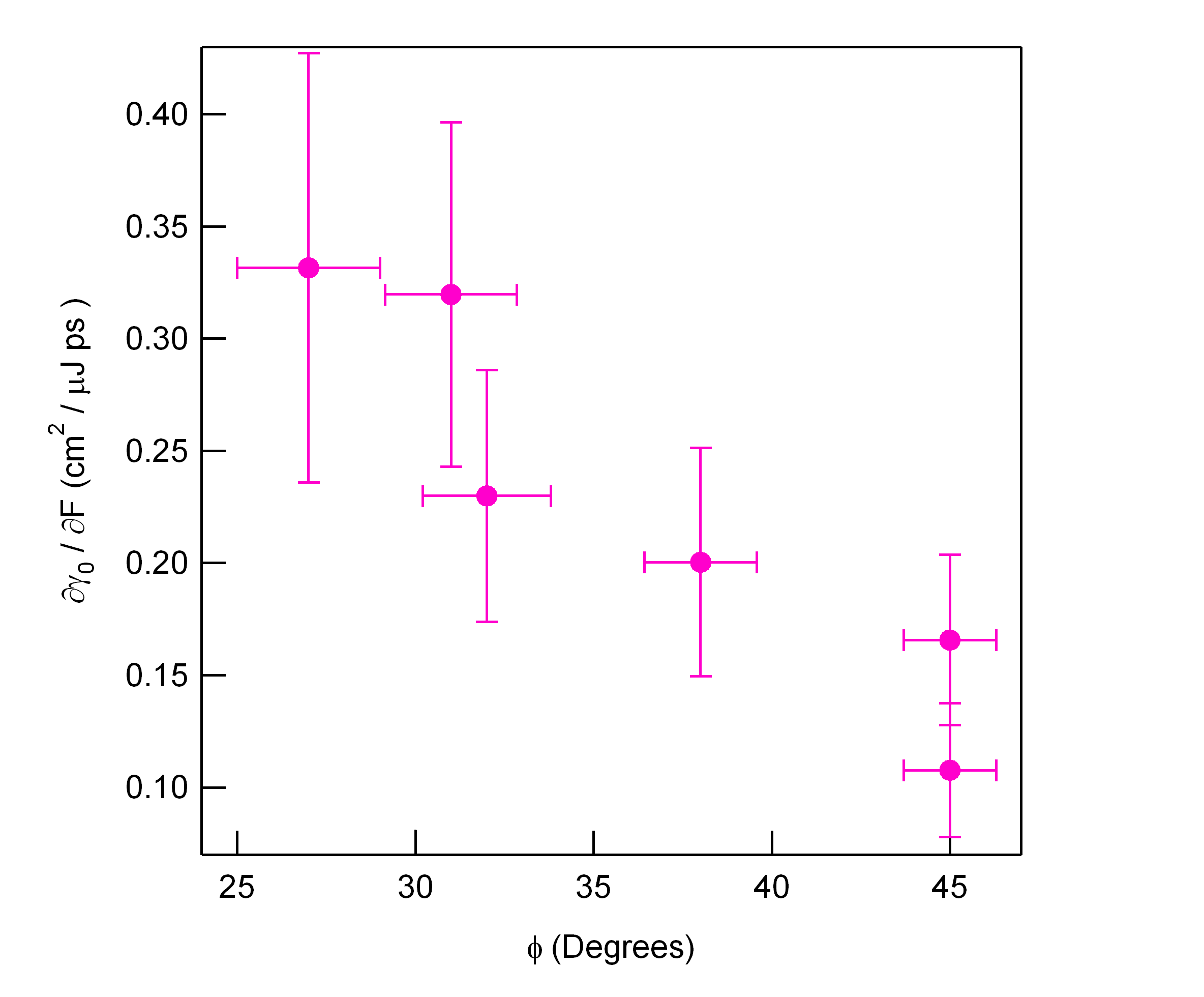}
\caption{\label{model} 
Initial rate of increase $\partial \gamma_0/\partial F$ as extracted from straight line fits to the data in Fig.\ 3D for fluence $F < 12$ \uJcm. The horizontal axis corresponds to the Fermi surface angle.
}
\end{figure}
It is clear that $\partial \gamma_0/\partial F$ increases with decreasing Fermi surface angle $\phi$, which means that the rate of recombination is enhanced as the quasiparticle momentum moves farther from the node.  

One potential scenario for the momentum dependence of  the recombination rates is that with increasing distance from the node the quasiparticle energy and momentum approach resonance with charge or spin density wave fluctuations to which the electrons are strongly coupled.
For example, a prominent neutron spin resonance is observed in \bisco\ along the (1,1) momentum vector\cite{Fong99}. Resonance between this mode and a quasiparticle pair would occur at a Fermi surface angle of about 12$^\circ$, leading to the prediction of a peak in $\partial \gamma_0 / \partial F$ at this Fermi surface angle.
We believe that demonstrating that recombination can be mapped using time-resolved ARPES and observing its strong momentum dependence will further stimulate development of pulsed sources that are capable of reaching all the relevant regions of momentum space.



\begin{scilastnote}
\item[] {\bf Acknowledgments:} We thank R.\ A.\ Kaindl, D.\ A.\ Siegel, S.\ D.\ Lounis, T.\ Miller, and R.\ Johnson for useful discussions. This work was supported by the Director, Office of Science, Office of Basic Energy Sciences, Materials Sciences and Engineering Division, of the U.S.\ Department of Energy under Contract No.\ DE-AC02-05CH11231.

\item[] {\bf{Supporting Online Material}}
\newline www.sciencemag.org
\newline Materials and Methods
\newline SOM Text
\newline Figs. S1 to S5
\newline References ({\it{30-35}}) 
\end{scilastnote}






\clearpage

\renewcommand{\theequation}{s\arabic{equation}}
\renewcommand{\thefigure}{S\arabic{figure}}
\setcounter{figure}{0}
\setcounter{equation}{0}

\section*{Supporting Online Material}
\subsection*{Materials and Methods}

The time-resolved ARPES experiment uses the same setup as that reported in Ref.\ \cite{Graf11}.
We induce a nonequilibrium state by exciting the sample with an infrared pump pulse ($h \nu = 1.48$ eV). 
This photon energy is sufficient to energize the system's quasiparticles, but does not eject them from the sample.
Shortly after photoexcitation, the sample is excited with a probe pulse in the ultraviolet (frequency-quadrupled from the pump using two BBO crystals, resulting in $h\nu = 5.9$ eV). 
The second pulse photoemits the electrons, and their energies and momenta are subsequently measured using a hemispherical analyzer with a 2D imaging detector (SPECS Phoibos 150).  
The total experimental energy resolution, including probe beam bandwidth, was measured to be 23 meV.  
The momentum resolution at a given photoelectron kinetic energies and emission angle was 0.003 \AA$^{-1}$.
Time resolution was achieved by a motorized translation stage which adjusts the relative pump and probe path lengths, and therefore the time delay between the two.  
The cross-correlation between the two pulses, measured using the response of polycrystalline gold 1 eV above the Fermi level ($E_F$), was about 300 fs.
The laser repetition rate was 543 KHz.
The typical pump pulse beam profile was 100 $\mu$m at full-width half-maximum (FWHM), and the typical probe pulse profile was 40 $\mu$m FWHM.
The sample was mounted on a 6-axis manipulator, cleaved and maintained at 18 K at a pressure below 5e-11 Torr, and oriented by taking ARPES Fermi surface maps.
ARPES spectra were corrected for detector non-linearity and obtained in a photocurrent regime where space charge effects were small.
Spectra at different delay times were also normalized to each other at high binding energy (100-200 meV), where the effect of photoexcitation is not detectable\cite{Graf11}, but where the overall intensity is seen to vary slightly with time due to variations in photocurrent. The typical correction, over the course of 20 ps, was by less than 2\% of the high binding energy intensity.

The time-resolved reflectivity experiment was performed using the same laser system as the ARPES experiment, also operating at a repetition rate of 543 kHz.  Both pump and probe pulses were infrared and were focused onto a 54 $\mu$m spot on the sample.  The pump delay was modulated using a rapid scan delay line operating at 20 Hz, with the pump amplitude modulated by a photo-elastic modulator (PEM) operating at 100 kHz.  The reflected probe beam was detected by a silicon photodiode, which was output to a lock-in amplifier referenced to the PEM modulation.

Measurements were performed on different cleaves of the same sample, comprising four data sets.  
The data corresponding to a Fermi surface angle of $\phi=27^\circ$ in Figs.\ 2, 3, and 4 are from the first cleave.  
The data corresponding to $\phi=32^\circ$, $\phi=38^\circ$, and $\phi=45^\circ$ (cut 5) in Figs.\ 2, 3, and 4 are from the second cleave. 
The data corresponding to $\phi=31^\circ$ and $\phi=45^\circ$ (cut 6) in Figs. 1, 3, and 4 are from the third cleave.
The time-resolved reflectivity data are from the fourth cleave.

\subsection*{ARPES Detector Non-Linearity Correction}

A common feature in even the best photoemission detectors is that the detector responds in a non-linear fashion to electron events\cite{Kay01,Mannella04,Seah99,Wicks09}. The effect is more pronounced with laser ARPES than synchrotron light because of the smaller background and higher electron flux. A typical curve showing the relationship between electron events and measured counts is displayed below in Fig.\ \ref{linearitycurve}.
\begin{figure}[htbp]\centering\includegraphics[width=\columnwidth]{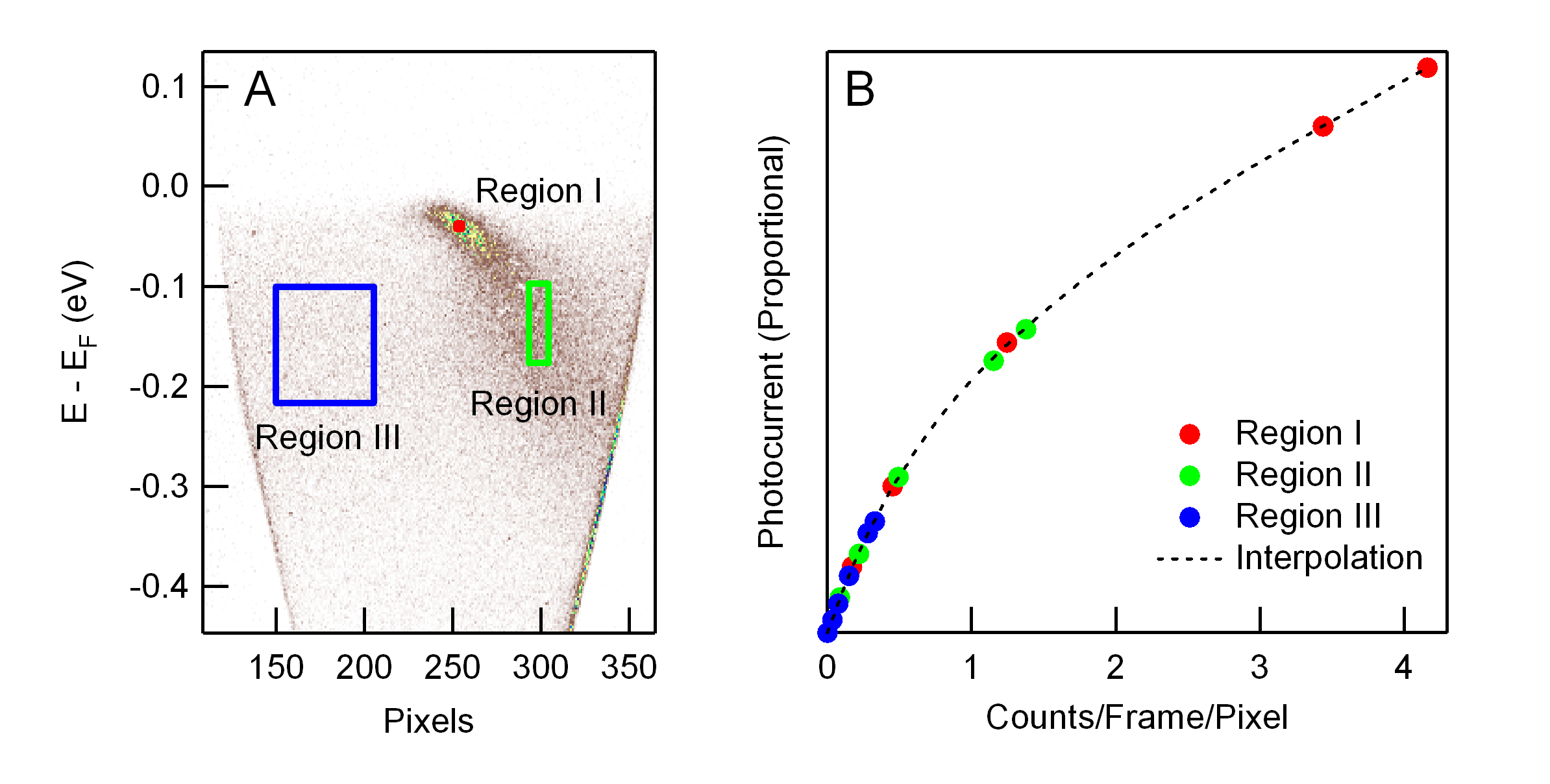}
\caption{\label{linearitycurve} Calibration method for detector non-linearities. (A) Typical raw ARPES image, used for calibrating detector non-linearity effects. Images were taken at several different photocurrents, and the detector response was measured in a high intensity region (Region I), moderate intensity region (Region II) and low intensity region (Region III). (B) Photocurrent vs.\ measured intensities for the 3 regions shown in (A). Responses from the three regions have been vertically scaled to match a single curve.}
\end{figure}

As can be seen from the figure, the response is linear for the lowest count rates ($< 0.4$ counts/frame/pixel) but deviates from this at moderate count rates. 
In order to adjust for these effects, we measured the sample using several different laser intensities before proceeding to the pump-probe experiment. 
Using a picoammeter that is attached to the sample, we were able to associate each laser intensity with a photocurrent. Photocurrent and laser power are proportional to each other\cite{Graf10}. By plotting photocurrent against the detector response in counts/frame/pixel, we then obtained the curve in Fig.\ \ref{linearitycurve}, and we applied this curve to the pump-probe data to remove the influence of non-linearity effects in the final analysis.
We note that the primary focus of the paper is the above-$E_F$ spectral weight, where the count rate is low and thus non-linearity corrections are small.

\subsection*{Superstructure}

All ARPES data were taken using cuts through the Fermi surface along the $\Gamma-Y$ direction in $k$-space in order to avoid complications with umklapp replica bands. The $(\sqrt{2}\times\sqrt{2})R45^\circ$ shadow band is beyond the $k$-space windows examined and has no influence on the data. While higher order umklapp replicas of the main band and shadow band can be observed when a sufficient volume of statistics is accumulated, such replica bands are orders of magnitude dimmer than the main band at this photon energy, and are not observed in these data sets.

Bilayer splitting in \bisco\ is visible at a photon energy of 5.9 eV, as seen in Fig.\ 1. 
In the off-nodal cut such splitting is visible in the form of the horizontal steak of intensity below the Fermi level for $k>k_F$. Bilayer bands are also present along the nodal direction, although at the node they effectively lie on top of one another.

\subsection*{Fluence Determination}
Assuming a Gaussian intensity profile, the total average fluence is given by
\begin{equation}
F_{ave} = (\textrm{Geometrical factor}) \times \frac{0.88 \, E_{pu}}{\textrm{FWHM}_{pu}^2+\textrm{FWHM}_{pr}^2}, \label{fleq}
\end{equation}
where $E_{pu}$ is the pump pulse energy, and $\textrm{FWHM}_{pu}$ and $\textrm{FWHM}_{pr}$ are the pump and probe full-width half-maximum intensity values.
In the experiments here reported, the pump and probe spot size were measured in two different ways: 
a) by using a flip mirror to divert the beam away from the chamber, toward a CCD beam profiler; 
and b) by using a pinhole mounted directly on the sample manipulator.  The two methods gave same value within 10\% of the beam diameter.
The geometrical factor arises from the fact that both pump and probe beams are incident on the sample from an oblique angle.
When the beam approaches a sample at an angle $\theta$ from normal incidence, it will spread into a spot that is larger by a factor of $1/\sin \theta$.  
The largest correction resulting from this effect and applied to our data corresponds to a 20\% adjustment downward in the fluence for the cut at $\phi=27^\circ$.

In order to verify the decay rate momentum dependence and definitively separate it from potential systematic geometrical effects, two different geometrical configurations were used.  
Data corresponding to cuts 1, 3, 4, and 5 in Fig.\ 3 were obtained by rotating the sample flip stage, which has the benefit of giving sharp, relatively symmetric cuts across the Fermi surface, but results in a spot size that changes with Fermi surface angle for a given pump power.
Data corresponding to cuts 2 and 6 were obtained by rotating the sample azimuth, where pump and probe spot sizes are identical between nodal and off-nodal cuts. 
The two types of measurements were in good agreement.

\subsection*{Error Bars}

Special care was taken to determine the accuracy of the fluence measurements.
The factors contributing to fluence error include variations in the power meter reading, losses from the viewport windows, pump-probe misalignment, pump spot size, and probe spot size. 
Of these, pump-probe misalignment is the most important cause of statistical uncertainty between one fluence measurement and the next in the ultrafast ARPES experiment. 
A misalignment of $\delta = 0.2 \times \textrm{FWHM}_{pu}$ for a pump-probe ratio of 2:1 can result in an average incident fluence that differs from the $\delta=0$ value by 8\%.
We determined the combined fluence error according to the formula,
\begin{equation}
\left( \frac{\sigma_{F}}{F} \right)^2 = \underbrace{\left( \frac{\sigma_{P}}{P} \right)^2}_{\textrm{power meter}}
+ \underbrace{0.03^2}_{\textrm{windows}} +  \underbrace{\left(\frac{A \delta}{\textrm{FWHM}_{pu}} \right)^2}_{\textrm{alignment}}
+ \underbrace{\frac{4 \, \textrm{FWHM}_{pu}^2 \, \sigma_{pu}^2 + 4 \, \textrm{FWHM}_{pr}^2 \, \sigma_{pr}^2}{ \left( \textrm{FWHM}_{pu}^2+\textrm{FWHM}_{pr}^2 \right)^2}}_{\textrm{spot sizes}},
\end{equation}
where $F$ is the fluence, $P$ is the laser power, $A$ is a proportionality factor, $\delta$ is the pump-probe spacial misalignment, and $\textrm{FWHM}_{pu}$ and $\textrm{FWHM}_{pr}$ are the respective pump and probe full-width half-maximum intensity values as in Eq.\ \ref{fleq}. The variables $\sigma$ refer to the standard error of their respective subscripts.

For the ultrafast ARPES measurements, we estimate 10\% fractional power meter uncertainty, 3\% fractional uncertainty due to window losses, 10\% fractional uncertainty from pump-probe alignment, 10\% uncertainty in the respective pump and probe FWHM values, and a typical pump-probe ratio of 2:1, which leads to a combined fractional fluence uncertainty of 22\%. 
For the ultrafast reflectivity measurements, the pump and probe had the same spot size, about 54 $\mu$m, and alignment was easier due to a smaller cryostat, comparable pump and probe spot sizes, and the fact that both beams were of the same frequency.
We estimate 10\% fractional power meter uncertainty, 3\% fractional uncertainty due to window losses, 5\% fractional uncertainty from pump-probe alignment, 10\% uncertainty in the respective pump and probe FWHM values, and a typical pump-probe ratio of 1:1, which leads to a combined fractional fluence uncertainty of 18\%. 

\begin{figure}[htbp]\centering\includegraphics[width=0.8\columnwidth]{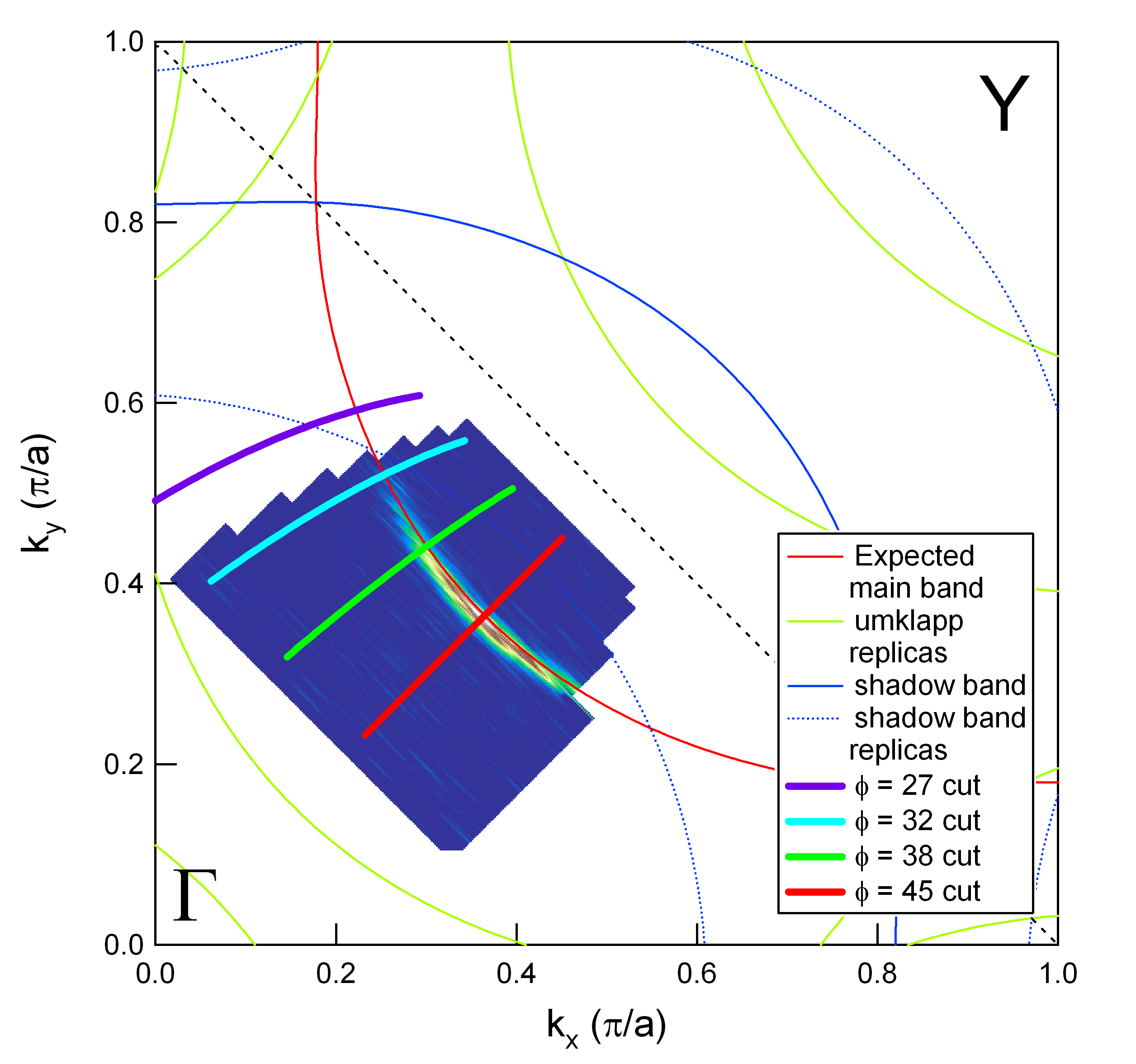}
\caption{\label{kmap} Low temperature constant energy map of \bisco\ at $E=E_F$. The expected main band, in red, is from a tight binding model\cite{Norman95}.}
\end{figure}

Horizontal error bars in Fig.\ 4 are based on potential skewing of the photoelectron exit angle from stray electric fields inside the chamber, based on a comparison between a tight-binding model for \bisco\cite{Norman95} and $k$-space maps that were acquired at $E=E_F$ at low temperature.
An analyzer work function $\Phi_W=4.35$ eV was found to produce an adequate match between the data and the expected main band.
We estimate the error in Fermi surface angle to be on the order of $\sigma_\phi \approx 2^\circ$ at $\phi=27^\circ$, and not more than $\sigma_\phi \approx 1.3^\circ$ in the nodal direction, which can be aligned more accurately because of the inherent symmetry of the band structure. An example map of the low temperature Fermi surface is displayed in Fig.\ \ref{kmap}.

Vertical error bars in Fig.\ 3 are based on fit standard deviations. Vertical error bars in Fig.\ 4 are based on the quadrature sum of the fit standard deviation from Fig.\ 3, and 22\% of the fit value, which incorporates the uncertainty in fluence.

\subsection*{Sample Heating}

\begin{figure}[htbp]\centering\includegraphics[width=0.8\columnwidth]{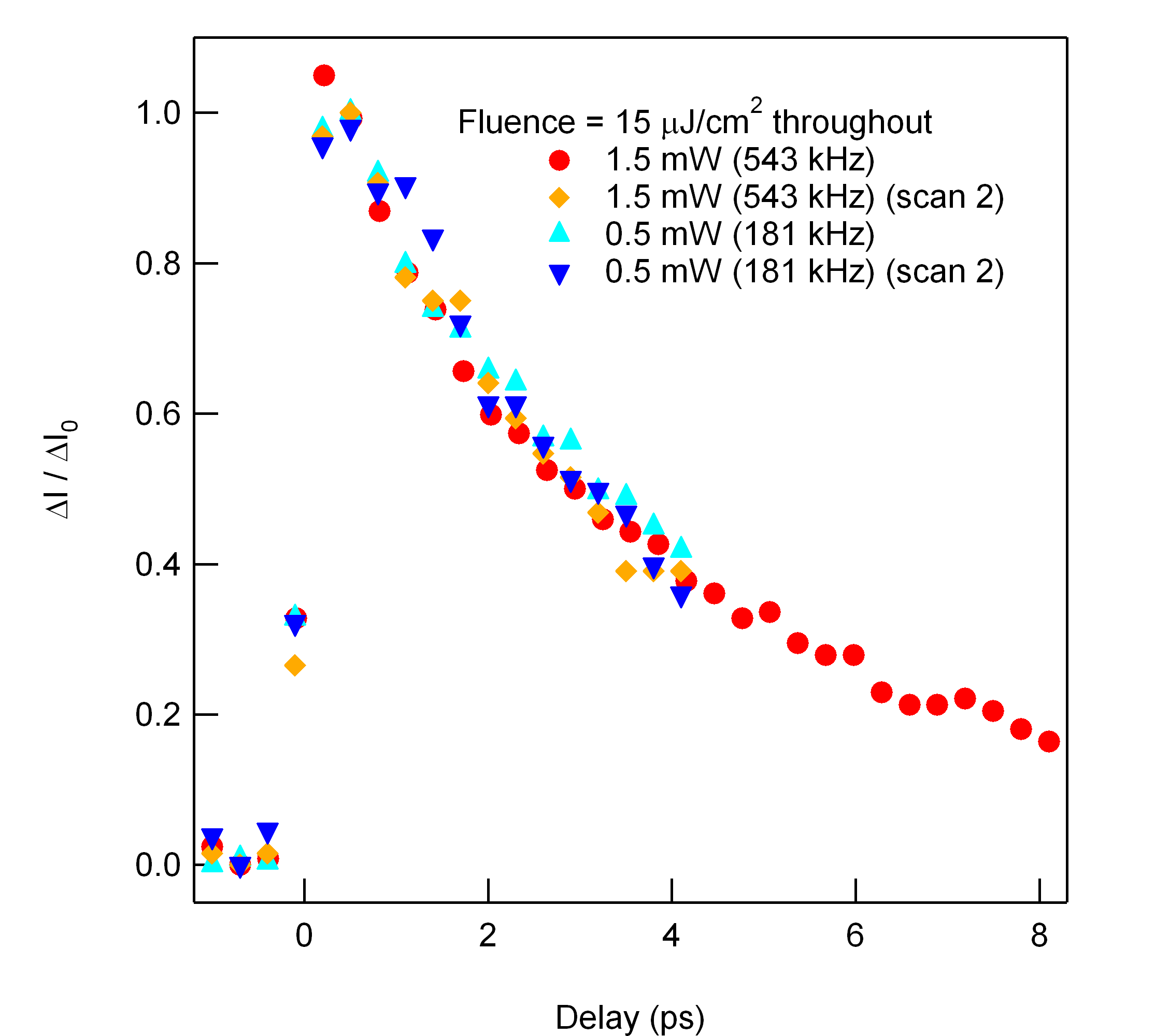}
\caption{\label{reprate} Dependence of spectral decay on laser repetition rate. Red circles and orange diamonds correspond to 543 kHz, which is the repetition ratio used for all data discussed in the main text. Blue and cyan triangles correspond to 181 kHz.}
\end{figure}

An important systematic effect to rule out is the possibility that the fluence dependence is not indicative of the incident pulse, but is rather the manifestation of a slight rise in equilibrium temperature caused by time-averaged heating due to the pump pulse.
We tested for this by reducing the repetition rate of the laser from 543 kHz to 181 kHz while maintaining a fluence of 15 \uJcm. 
The residual heating was thus equivalent to that caused by a fluence of 5 \uJcm\ at 543 kHz. 
Fig.\ \ref{reprate} shows a comparison of decay rates as measured using the two settings.
There is no discernible difference in the decay curves, which verifies that heating cannot be responsible for the observed fluence dependence.

\subsection*{Superconducting Gap Measurements}

The gap values in Fig.\ 2 were extracted according to the methodology of Norman et al.\cite{Norman98a}, which is a commonly accepted procedure in the field. We fit symmetrized EDCs at the Fermi momentum to the convolution of a Gaussian (FWHM = 20 meV) and the function
\begin{equation}
A(\omega) = \textrm{Const.} \times \frac{\Gamma}{\left( \omega - \Delta_k^2/\omega \right)^2 + \Gamma^2}. \label{gapfunc}
\end{equation}
This amounts to a self energy ansatz of $\Sigma(\omega) = \Delta_k^2/\omega - i \Gamma$, where a phenomenological constant $-i\Gamma$ has been added to the BCS self energy correction in order to account for finite lifetime effects.

\begin{figure}[htbp]\centering\includegraphics[width=\columnwidth]{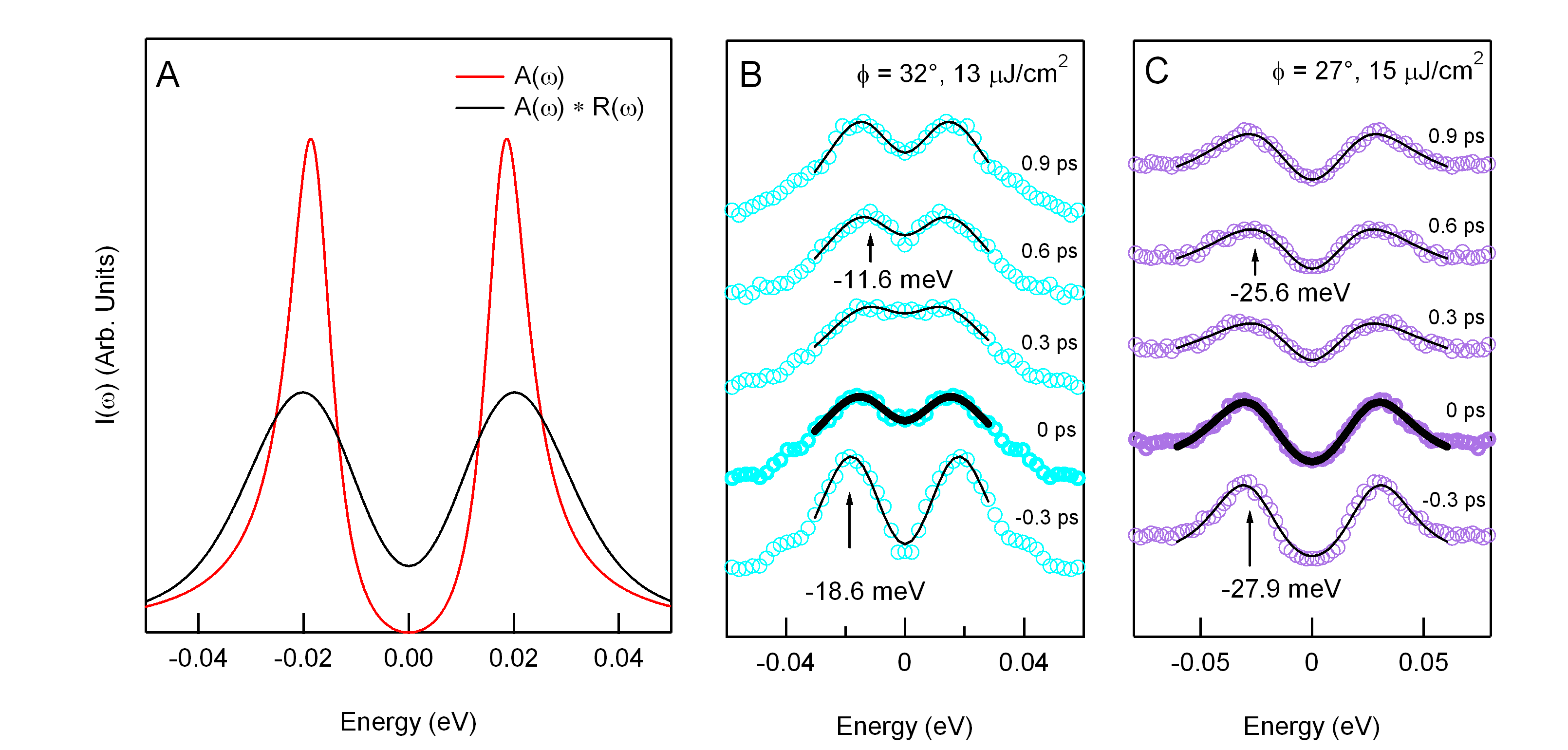}
\caption{\label{gapmodel2}(A) Graph of Eq.\ \ref{gapfunc} and its convolution with a Gaussian resolution function $R(\omega)$ of FWHM = 20 meV. (B) Symmetrized EDCs and EDC fits at $k_F$ for a momentum cut at $\phi=32^\circ$. (C) Symmetrized EDCs and EDC fits at $k_F$ for a momentum cut at $\phi=27^\circ$.}
\end{figure}

A graph of Eq.\ \ref{gapfunc} is shown in Fig.\ \ref{gapmodel2}A along with the resolution-convolved form that was used to fit the data. 
The relevant parameters are $\Delta_k=19$ meV and $\Gamma = 10$ meV, which produce a curve that adequately matches the top symmetrized EDC in Fig.\ 2A of the main text.
Because Eq.\ \ref{gapfunc} has asymmetric quasiparticle peaks, the fitted gap size is slightly smaller than the maximum values of the resolution-convolved peaks would suggest.
In Fig.\ \ref{gapmodel2}B-C we show a magnification of the symmetrized EDCs from Fig.\ 2 of the main text corresponding to the highest pump fluence values. To complement the gap values extracted from the fitting analysis, we provide a comparison to gap values extracted from explicit measurements of peak locations in the symmetrized EDCs.
At $\phi=32^\circ$ the equilibrium EDC peak at $t=-0.3$ ps is 18.6 meV. At $t=0.6$ ps, this value is 11.6 meV, a reduction of 38\%. At $\phi=27^\circ$ the equilibrium EDC peak shrinks from 27.9 meV at equilibrium ($t=-0.3$ ps) to 25.6 meV at $t=0.6$ ps, an 8\% reduction.
These changes are in qualitative agreement with the change in gap size as measured using the fit, which is reduced by 42\% at $\phi=32^\circ$, and 19\% at $\phi=27^\circ$. In both cases the gap at $\phi=32^\circ$ is more strongly affected by the pump pulse than the gap at $\phi=27^\circ$. 
The fit yields a larger change in gap size than the peak-to-peak analysis, which may indicate that the finite energy resolution and data noise impose limits on the effectiveness of the peak-to-peak analysis, or it may reflect limitations of the fitting algorithm. Fig.\ 2 in the main text displays fitted gap values rather than results of a peak-to-peak analysis because the fit makes use of more data points than the peak-to-peak analysis, because it adequately captures asymmetries in the EDC lineshapes that would not be accounted for in simpler fitting algorithms, and because it rests on a firmer theoretical foundation than the simpler peak-based extraction method.

\begin{figure}[htbp]\centering\includegraphics[width=\columnwidth]{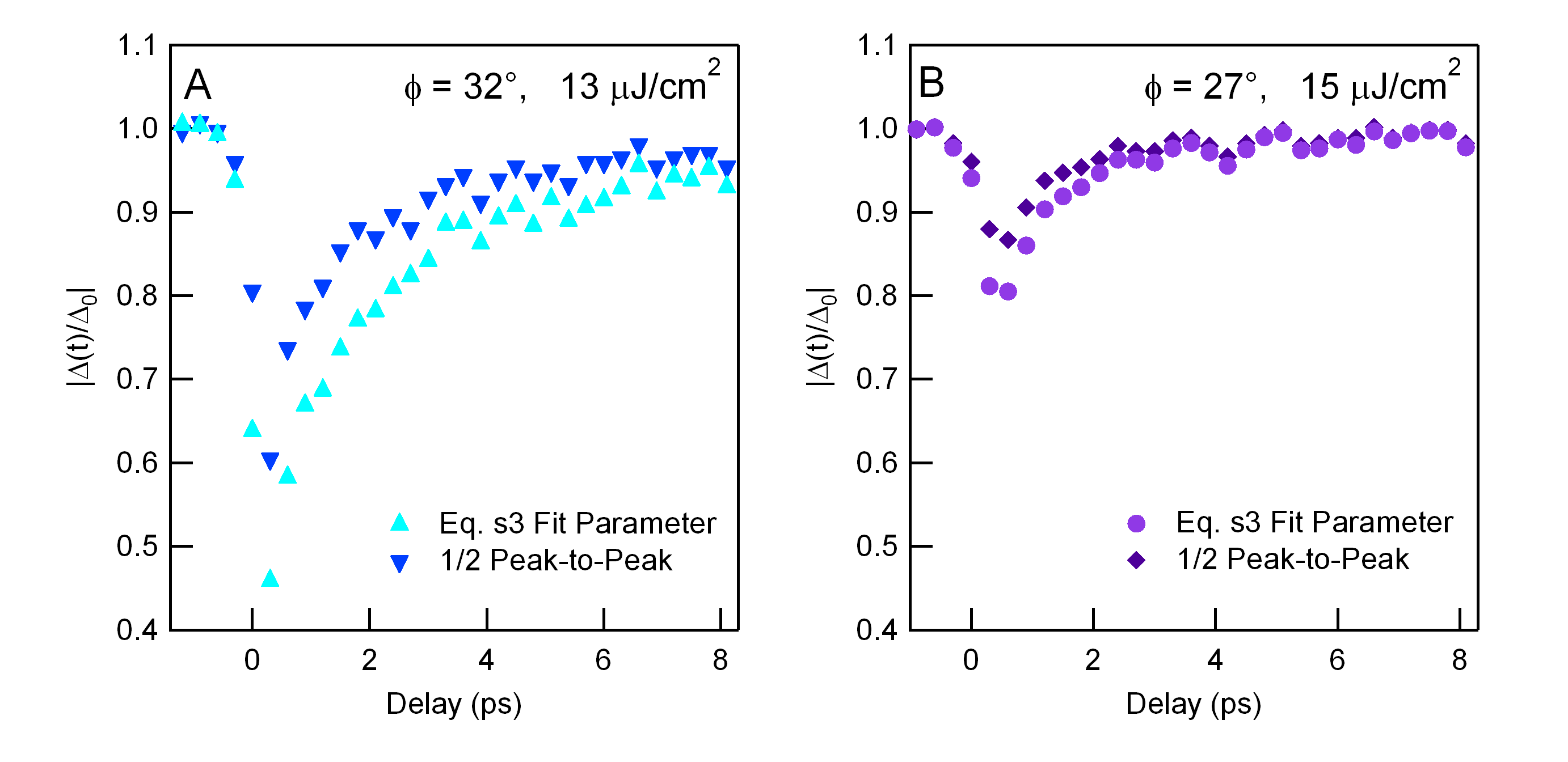}
\caption{\label{gapmodel3}(A) Normalized change in the equilibrium gap size for a cut at $\phi=32^\circ$. Cyan triangles correspond to the fitted gap parameter from Eq.\ \ref{gapfunc}. Blue inverted triangles correspond to 1/2 of the peak-to-peak distance of the fit curves in Fig.\ \ref{gapmodel2}B. (B) Normalized change in the equilibrium gap size for a cut at $\phi=27^\circ$. Light purple circles correspond to the fitted gap parameter from Eq.\ \ref{gapfunc}. Dark purple diamonds correspond to 1/2 of the peak-to-peak distance of the fit curves in Fig.\ \ref{gapmodel2}C. }
\end{figure}

Further comparison between the extracted gap and a simpler peak-to-peak analysis is shown in Fig.\ \ref{gapmodel3}. 
The lighter triangles and circles correspond to the fitted gap parameter from Eq.\ \ref{gapfunc}. The darker triangles and diamonds correspond to 1/2 peak-to-peak distances from the fit curves in Fig.\ \ref{gapmodel2}B-C. As is the case in comparison to the raw data above, the fit parameter yields a larger change in gap size than the peak-to-peak analysis, but the cut at $\phi=32^\circ$ is more strongly affected by photoexcitation than the cut at $\phi=27^\circ$ under any analysis scheme.

\subsection*{Recombination Model}
\paragraph{Comparison of Reflectivity and ARPES Decay Rates.}

While the Rothwarf-Taylor model is a widely accepted framework for describing relaxation processes in the cuprates, there are competing views as to whether or not cuprate dynamics lie within the boson bottleneck limit of these equations.
Under bottleneck conditions, hot bosons regenerate quasiparticle pairs much faster than they anharmonically decay or diffuse into the bulk of the sample. In the absence of a bottleneck, the opposite is true. Nevertheless, in the low temperature and low fluence limit both viewpoints reduce to the same functional form, which is the following at zero temperature,
\begin{align}
  \dot{n}_{ex} = -R \, n_{ex}^2 \quad &\Rightarrow \quad
  n_{ex}^{-1}(t) = n_{ex}^{-1}(0) + R \, t, \label{rt} \\
\intertext{or finite temperature,}
  \dot{n}_{ex} = -R \left( n_{ex}^2 + 2 n_{ex} n_T \right) \quad &\Rightarrow \quad
  n_{ex}^{-1}(t) = \left( \frac{1}{2 n_T} + \frac{1}{n_{ex}(0)} \right) e^{2 n_T R t} - \frac{1}{2 n_T}. \label{rtfinite}
\end{align}
Here $n_{ex}(t)$ is the population of photoexcited quasiparticles, $n_T$ is the population of thermal quasiparticles, and $R$ is the intrinsic Cooper pair recombination rate in the bottleneck-free limit, or a renormalized Cooper pair recombination rate in the bottleneck limit.
A more complete explanation of how these equations follow from the full Rothwarf-Taylor model is described elsewhere\cite{Gedik04,Kabanov05}. 

In Fig.\ 3 of the main text, ARPES and reflectivity quasiparticle response curves are compared 
at short times by using a fit to an exponential function $f(t)=\Delta I_0 e^{-r_0(t-t_0)}\Theta(t-t_0)$.
The short-time limits of Eqs.\ \ref{rt} and \ref{rtfinite} are respectively 
\begin{align}
\frac{n_{ex}(t)}{n_{ex}(0)} &= 1 - R \, n_{ex}(0) \, t + O(t^2) \\
\intertext{and}
\frac{n_{ex}(t)}{n_{ex}(0)} &= 1 - R[n_{ex}(0) + 2 n_T] t + O(t^2).
\end{align}

In the case of ARPES, 6 eV photons induce photoelectrons with a mean free path of 5 nm.
This is well below the penetration depth of the 1.48 eV pump beam (100 nm), which means that only the surface of the material is probed. Furthermore, the initial photoexcited population $n_{ex}(0)$ vastly outweighs the thermal population $n_T$ in the majority of cases.
To a good approximation, then, the decay rate fits displayed in Fig.\ 3 correspond to the product $R \, n_{ex}(z=0,t=0)$. 

In contrast, the optical reflectivity probe penetrates the same distance into the sample as the pump, and therefore necessitates a more intricate analysis of spacial variation in $n_{ex}(z,t)$.
The problem has been addressed by Gedik et al.\cite{Gedik04}, who propose that the optical response should be modeled by inserting the solution of Eq.\ \ref{rt} into a weighted integral as follows:
\begin{eqnarray}
\Delta R(t) &=&  \frac{2\Delta R(0)  \alpha}{n_0} \int_0^\infty dz \, e^{-\alpha z} n_{ex}(z,t), \qquad n_{ex}(z,t) \equiv \frac{n_{ex}(0,0)e^{-\alpha z}}{1 + R n_{ex}(0,0)e^{-\alpha z} t} \\
&& \nonumber \\
&=& \frac{2\Delta R(0)}{\gamma_0 t} \left[ 1 - \frac{\ln(1+\gamma_0 t)}{\gamma_0 t} \right]. \label{gedik}
\end{eqnarray}
The constant $\gamma_0 \equiv R \, n_{ex}(0,0)$ corresponds to the initial decay rate of quasiparticles at the surface. Finite temperature effects are ignored because $n_T \ll n_{ex}(0)$ (as noted above). In the short-time limit, the decay rate of this revised equation differs by a factor of 2/3 from its counterpart at the sample surface:
\begin{equation}
\frac{\Delta R(t)}{\Delta R(0)} = 1 - \frac{2}{3} \, \gamma_0 \, t + O(t^2).
\end{equation}
This is the most important difference between the physical interpretation of ARPES and optical decay rates in the short-time limit. 

\paragraph{Momentum-Dependent Recombination.}
In order to allow for momentum-dependent recombination processes, Eq.\ \ref{rtfinite} may be recast as follows:
\begin{equation}
\dot{n}_k = - n_k \int R_{kk'} \, n_{k'} \, d^2k' - n_k \int R_{kk'}\, n_{k'T} \, d^2k' - n_{kT} \int R_{kk'}\, n_{k'} \, d^2k' 
\end{equation}
where $n_k$ is now the population of photoexcited quasiparticles at a given momentum $k$, $R_{kk'}$ is a recombination function related to the formation of Cooper pairs, and $n_{kT}$ is the momentum-dependent population of thermal quasiparticles. The first term on the right corresponds to photoexcited quasiparticles recombining with other photoexcited quasiparticles. The second term corresponds to photoexcited quasiparticles at $k$ recombining with thermal quasiparticles in other parts of the Brillouin zone. The final term corresponds to photoexcited quasiparticles in other parts of the Brillouin zone recombining with thermal quasiparticles at $k$. The integral in the second term is a constant, so the term does not contribute to any density or fluence dependence. The third term vanishes for all momenta except the nodal direction, and along the node it is small relative to the first term on the right for all but the lowest excitation densities. At short time, then, momentum-dependent recombination dynamics are dominated by the first term on the right.

In general, $n_k$ is a function of the fluence $F$, and can thus be written as $n_k(F)$. Because $n_k$ is the photoexcited quasiparticle population $n_k(0)=0$. As a result, for small fluence $n_k(F)$ can be expanded to linear order in $F$ as $n_k \approx \alpha_{k'} F$. Making this substitution to the integrand in the first term results in 
\begin{equation}
\dot{n}_k \approx - n_k \int R_{kk'}\, n_{k'} \, d^2k' \approx - n_k \int R_{kk'}\, \alpha_{k'} \, F \, d^2k'  
\end{equation}
After pulling $F$ outside of the integral and taking the derivative, the approximation finally reduces to
\begin{equation}
\gamma_{k0} \approx - F \int R_{kk'}\, \alpha_{k'} \, d^2k' \qquad \Rightarrow \qquad \frac{\partial \gamma_{k0}}{\partial F} \approx \int R_{kk'}\, \alpha_{k'} \, d^2k',
\end{equation}
where $\gamma_{k0} \equiv \dot{n}_{k0} / n_{k0}$ is the momentum-dependent initial decay rate plotted in Fig.\ 3.

\end{document}